# A viewpoint from dissipative dynamics on diffusion-controlled directional solidification

Fengyi Yu

CAS Key Laboratory of Mechanical Behavior and Design of Materials, Department of Modern Mechanics, University of Science and Technology of China, Hefei 230026, China

E-mail address: fengyi.yu.90@gmail.com

**Abstract**

The existing theoretical analyses of solidification dynamics lack the insights of historical relevance and transport processes in the whole system. Through the phase-field model, this paper investigates the evolution in the whole domain during entire directional solidification. First, the evolution of characteristic parameters is obtained, including the solute concentration ahead of interface and tip velocity, demonstrating the dissipative features of solidification. Second, by adjusting the diffusion coefficient $D_L$, the dissipation at the interface can be altered. With different $D_L$, different stages during directional solidification are investigated, including planar growth and instability, dendrite growth, and steady-state growth. The results indicate the important role of solute diffusion in alloy solidification. From the viewpoint of the whole domain, smaller $D_L$ corresponds to a higher degree of dissipation, forming more interfaces during solidification. Moreover, the competitive influences of tip curvature and velocity are also because of the dissipation, resulting from the friction of atoms.

# 1. Introduction

The mechanical properties of as-solidified components are dominated by the solidification structures. The precise control of the structures requires a deep understanding of solidification dynamics. The dynamics is determined by the competition between the transport of heat and mass and the inherent length scale of the material induced by interfacial energy [1]. In this spirit, researchers developed various theoretical models to describe the dynamics at different stages of directional solidification [2-4]. For early-stage growth, including the planar growth and instability, the descriptions go through the Constitutional Supercooling (CS) theory [5], the Mullins-Sekerka (MS) analysis [6], to the Warren-Langer (WL) model [7]. After planar instability, solidification turns into dendrite growth. During dendrite growth, two stages are distinguished due to the different features. The first is unsteady growth; the velocity and shape of the dendrite tip vary with time. The second is steady-state growth; the velocity and shape are constant. The competitive influences of tip radius $\rho$ and velocity $V_I$ determine the dendrite growth [8,9]. Through the characteristic parameters $\rho$ and $V_I$, these theoretical insights express the relations between the dynamics and the solidification parameters: temperature gradient G and pulling speed $V_P$ [10].

However, the existing analyses lack the insights of historical relevance and the transport processes in the whole system. Specifically, solidification structures are formed out of equilibrium in an open system, i.e., a dissipative system [11]. Unlike equilibrium systems, dissipative systems develop patterns different from the initial states. Even under the same solidification conditions, different initial conditions make the final structures quite different. That is, solidification is a history-dependent process. The investigation during the entire process and the viewpoints from dissipative dynamics can provide new perspectives on solidification evolution.

Dissipative models always appear as coarse-grained models describing the averaged behavior of complex systems. Mean field models of thermodynamic origin are always dissipative. Generally, they are derived from energy and mass conservation laws, but contain sources that destroy conservation of the respective integrals [11]. As one of them, the Phase-Field (PF) model has solid physical foundations, also with high numerical accuracy [12-14]. Specifically, the physics underlying the construction of the PF model includes the insights of thermodynamics, the dynamics of transport, and the interfacial anisotropy [15,16]. When solving the governing equations, using an additional scalar field to implicitly represent the interface by one of its level sets, the PF method avoids the shape error caused by tracking interface. It can capture complex interfacial morphologies and characteristic parameters at the interface with high accuracy. With the significant advantages, the PF method has been increasingly applied to investigating solidification dynamics, including the planar-to-cellular transition [17,18], the selection of growth direction [19,20], and competitive growth [21,22], etc. The consistencies with the experimental observations demonstrate the accuracy of the PF method, ensuring that the PF model is suitable for investigating the dissipative dynamics of solidification.

Through the quantitative PF model, this paper investigates the evolution in the whole domain during entire directional solidification. First, the evolution of the characteristic parameters is discussed in detail, showing the dissipative features. Second, by adjusting the diffusion coefficient $D_L$, the dissipation of the system can be altered. With different $D_L$, different stages in directional solidification are investigated, including planar growth and instability, dendrite growth, and steady-state growth. Viewpoints from dissipative dynamics on diffusion-controlled directional solidification are given out.

## 2. Results and discussion

In this section, varying solidification parameters are adopted in the simulations. The thermal gradient G is constant at 100K/mm, while the pulling speed $V_P$ increases from 0 to a fixed value of 300μm/s, for which the increase time is 0.5s, shown in Fig. 1(a). The intersection angle between the TGD and PCO of the crystal is 30°. The domain of the PF simulation is 2400×2400 grids, corresponding to 288.0μm×288.0μm in the real unit. It takes about 30 hours using 40 cores to finish one job.

### 2.1 The dynamical evolution in a dissipative system

As mentioned before, solidification patterns are dissipative structures formed out of equilibrium in an open system. The dissipative systems develop structures they do not have when first formed. Hence, the dynamical evolution of the characteristic parameters with time should be described clearly to represent the dissipative features better.

Fig. 1 shows the evolution of the characteristic parameters, including the concentration at interface $c_0$ and the instantaneous velocity of interface $V_I$. In Fig. 1(a), the curves from the WL model and PF model show good consistencies before the crossover time (t=0.67s) of the planar instability, validating the accuracy of the PF model. After the crossover time, the curves from the WL model and PF model differ from each other. Since the parameters obtained from the WL model are based on the planar interface, after the planar instability, the interfacial morphologies lead to the differences between these two models. That is, the differences do not mean the contradictions between these two models. Since the PF model can represent complex interfacial morphologies, the following discussion of the characteristic parameters at the interface is based on the PF results.

Meanwhile, the sharp interface model of alloy solidification could describe the evolution of $c_0$ and $V_I$ qualitatively and describe their features intuitively. Equations (1)-(2) express the one-sided sharp interface model of alloy solidification, where $\kappa$ is the interfacial curvature and $\dot{T}$ is the cooling rate (=$G*V_P$). $\partial_n c|^+$ is the gradient of concentration at the liquid side of the interface. Since $\partial_n c|^+ <0$ in equations (2), to make the discussion intuitive, we set $|\partial_n c|=|\partial_n c|^+|$.

$$c_0 = \frac{c_\infty}{k} - \frac{\kappa \cdot \Gamma + G(z - V_P t)}{|m|} = \frac{c_\infty}{k} - \frac{\kappa \cdot \Gamma + Gz + \dot{T}t}{|m|} \tag{1}$$

$$V_I = \left(-D_L \cdot \partial_n c|^+\right)/\left[(1-k)c_0\right] = D_L \cdot |\partial_n c|/\left[(1-k)c_0\right] \tag{2}$$

According to the PF results and the sharp interface model, the dynamical evolution during directional solidification is the following. Fig. 1 shows the evolution of concentration $c_0$ and velocity $V_I$ during the entire solidification process. The corresponding morphological evolution at the different stages is shown in Fig. 2. At the initial stage, in Fig. 1(a1), both $c_0$ and $V_I$ increase. As time goes on, they reach the peaks and then decrease, shown in Fig. 1(a2). Through a period of oscillation, they turn into a steady state, shown in Fig. 1(a3). The detailed evolution at each stage is following:

At the planar growth stage, the accumulation of solute ahead of the interface makes $c_0$ increase. At this stage, the interfacial curvature is zero. As expressed in equation (1), both $V_P$ and $t$ increase with time. Hence, $c_0$ increases with time exponentially. Meanwhile, the increase of $c_0$ means $|\partial_n c|$ increases, in equation (2), making $V_I$ increase with time. The evolution of $c_0$ and $V_I$ at the planar growth stage is shown in Fig. 1(a1) and the corresponding morphological evolution is shown in Fig. 2(a).

As solidification goes on, the planar instability occurs, represented by the sharp increment of the $V_I$ curves in Fig. 1(a1) and (a2). The morphological evolution is shown in Fig. 2(a). On the one hand, when instability occurs, due to mass conservation, the concentration $c_0$ still increases after the crossover time. On the other hand, the cellular appear after the instability. Due to the curvature of the cellular, rather than diffusing only along the pulling direction of the planar interface, the solute can diffuse along multiple directions from the cellular tip to the liquid, resulting in the decrease of $c_0$ (see equation (1)). That is, after the planar instability, the concentration ahead of the interface first increases and then decreases, shown by the limited increment of the $c_0$ curve after the crossover time in Fig. 1(a2). For the velocity $V_I$, equation (2) shows the competitive influences of $|\partial_n c|$ and $c_0$ depend on their specific values. $|\partial_n c|$ is determined by $V_I$ (by adjusting the diffusion length) and $c_0$ (by adjusting the difference between $c_0$ and $c_\infty$), synergistically. It is evident that $|\partial_n c|$ has positive relations with both $V_I$ and $c_0$. After the planar instability, the increment of $c_0$ is very limited. By contrast, due to the increase of $c_0$ and the curvature of the cellular, the $|\partial_n c|$ increase greatly. Hence, the velocity $V_I$ increases greatly, shown by the sharp increment of the $V_I$ curve after the crossover time in Fig. 1(a2). After the peak of the $c_0$ curve, $c_0$ starts to decrease, and the velocity $V_I$ increase further (see equation (2)). As time goes on, the further decrease of $c_0$ makes $|\partial_n c|$ decrease. As a result, $V_I$ reaches the maximum value and then decreases. With the decrease of $V_I$, the interfacial curvature $\kappa$ decreases, making $c_0$ increase (see equation (1)). The increase of $c_0$ brings the increase of $|\partial_n c|$, making the velocity $V_I$ increase again. Then the increase of $V_I$ brings the increase of $\kappa$, making $c_0$ decrease again. The cycle goes on and on till the steady-state stage, shown by the oscillating curves (from t=1.0s to t=2.2s) in Fig. 1(a2) and the morphological evolution in Fig. 2(b)-(c).

As time goes further, solidification turns into steady-state growth, shown by the stable curves (after t=2.2s) in Fig. 1(a3) and the morphological evolution in Fig. 2(d). At this stage, the dissipative solidification structures achieve a quasi-steady state after a period of self-organization. Hence, both $c_0$ and $V_I$ turn into a steady state.

In conclusion, the consistencies at the planar stage between the WL model and PF model demonstrate the accuracy of the PF simulations. After the planar instability, the interfacial effects make the evolution of the concentration $c_0$ and tip velocity $V_I$ complex. As time goes on, $c_0$ and $V_I$ reach the peaks successively and then decrease. Subsequently, both the $c_0$ curve and $V_I$ curve show oscillatory behaviors. Finally, they turn into a steady state. The evolution of the characteristic parameters illustrates the dissipative features during directional solidification.

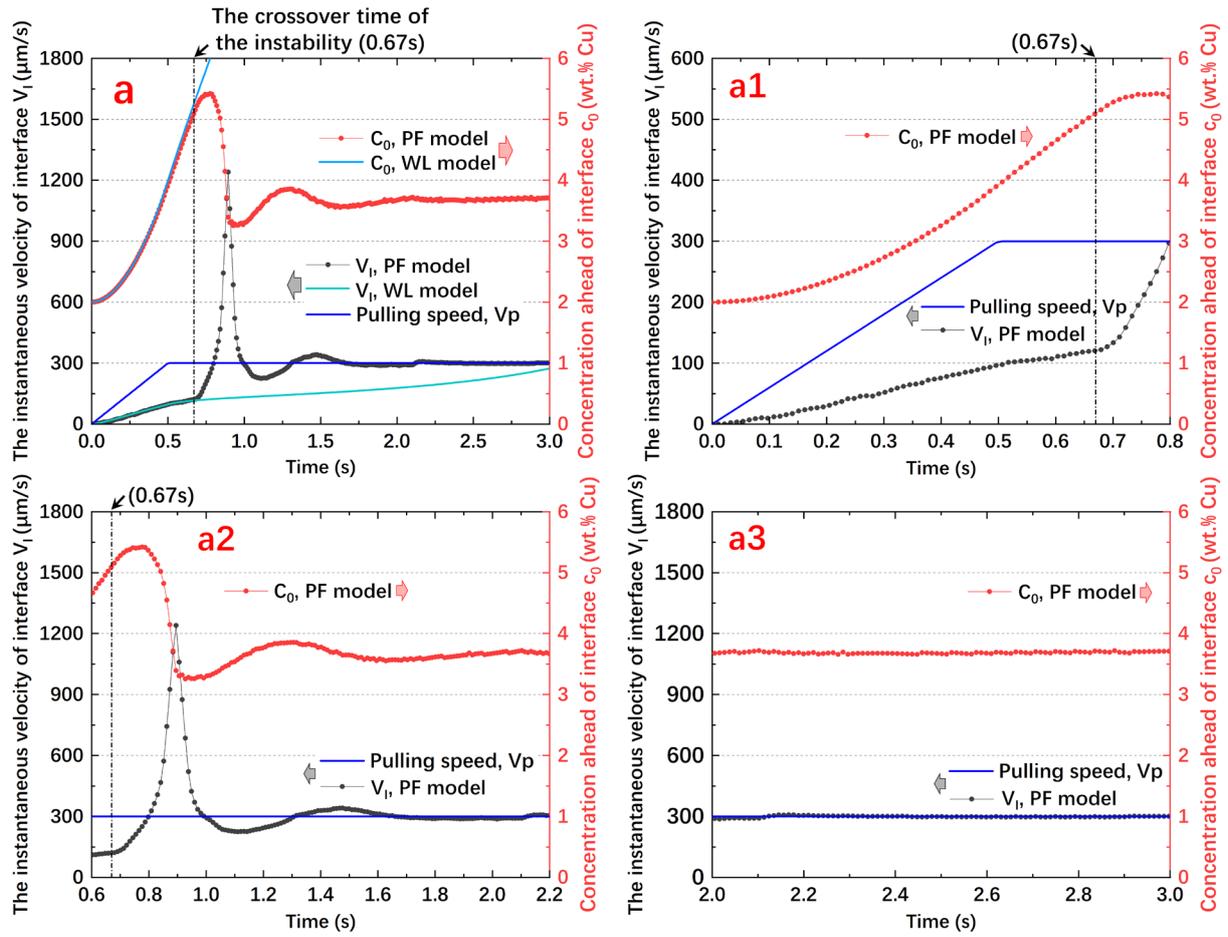

Fig. 1. The evolution of the characteristic parameters with time during the entire directional solidification: (a) the concentration ahead of the interface $c_0$ and the instantaneous velocity of the interface $V_I$; (a1), (a2), and (a3) are the enlarged versions of (a).

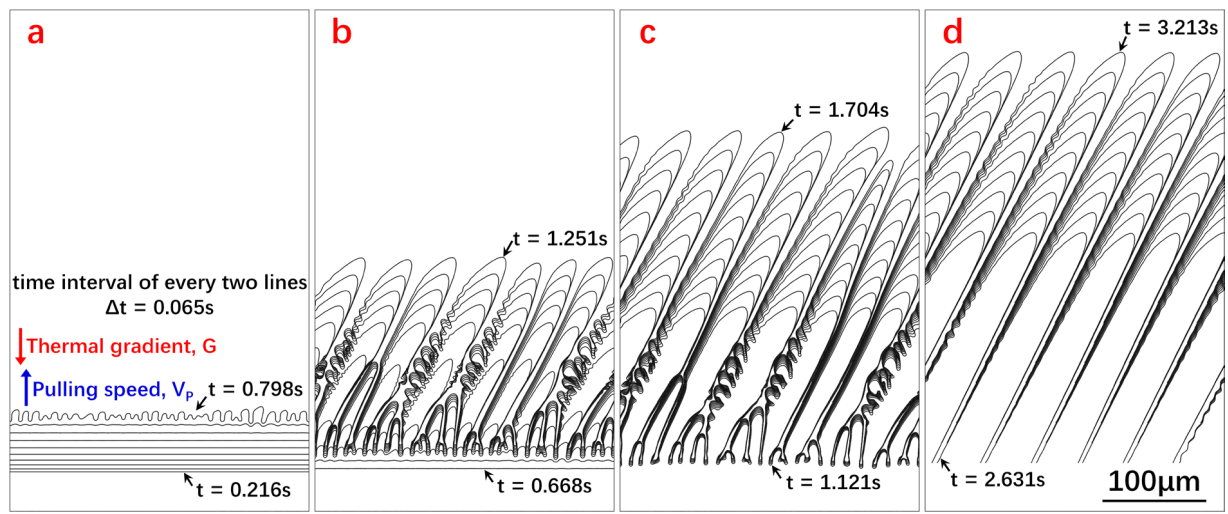

Fig. 2. The morphological evolution at different stages of the directional solidification: (a) the planar growth and instability; (b)-(c) the planar-to-cellular and dendrite growth (oscillation); (d) the steady-state growth.

## 2.2 The role of solute diffusion in a dissipative system

The previous section describes the dissipative features of solidification. The cause of dissipation and the factors affecting dissipation should be affirmed. According to statistical mechanics, dissipation results from the friction of atoms or molecules. Friction (dissipation) means the tendency of systems to move toward equilibrium when different degrees of freedom are allowed to interact with each other [23]. For alloy solidification, dissipation is the tendency of supersaturation to equilibrium. The diffusion coefficient $D_L$ and dissipation satisfy the relation

$$D_L = \frac{k_B T}{\gamma_0} \tag{3}$$

where $\gamma_0$ is the friction coefficient (dissipation). Equation (3) shows the coefficient $D_L$ has a negative relation with the degree of dissipation. That is, by adjusting the value of $D_L$, the dissipation of the system can be altered.

In this section, to represent different degrees of dissipation, different diffusion coefficients $D_L$ are used in the PF simulations. $D_L$ are set to be $2.0\times10^{-9}$, $3.0\times10^{-9}$, $4.0\times10^{-9}$ and $6.0\times10^{-9}$ $m^2/s$, respectively. It should be noted, the variation of $D_L$ here is just for parametric study, aiming to reveal the role of solute diffusion in morphological evolution. In the real physical world, $D_L$ cannot vary within such a great range.

With different $D_L$, the evolution of the concentration at the interface $c_0$ and tip velocity $V_I$ is shown in Fig. 3. On the one hand, the tendencies of $c_0$ and $V_I$ curves show similar characteristics, including the sharp increments of the $V_I$ curves, the peaks of the $c_0$ and $V_I$ curves, the oscillation of the curves, and the stable state of the curves. On the other hand, due to the different $D_L$, the degrees of dissipation are different during the evolution. Hence, the quantitative features of the evolution show differences, as follows.

### 2.2.1 The planar growth and instability

At the planar growth stage, the accumulation of solute ahead of the interface makes $c_0$ increase. At the same moment before the planar instability, in Fig. 3(a1), $c_0$ decreases from $2.0\times10^{-9}$ to $6.0\times10^{-9}$. For the velocity $V_I$, according to equation (2), $V_I$ has a positive relation with $D_L$ while having a negative relation with $c_0$. The larger $D_L$ and smaller $c_0$ correspond to the larger $V_I$. At the same moment before the instability, $V_I$ increases from $2.0\times10^{-9}$ to $6.0\times10^{-9}$, shown in Fig. 3(b1).

As solidification goes on, the planar instability appears, represented by the sharp increments of the $V_I$ curves in Fig. 3(b1)-(b2). To validate the PF results, the calculations of the theoretical model are performed, using the same parameters. By solving equations (10)-(12), the increase rate $\sigma_\omega$ and perturbation amplitude $A_\omega$ can be obtained, shown in Fig. 4. The time when $\sigma_\omega$ becomes positive represents the critical time of the marginal stability, shown by the red curves in Fig. 4(a1)-(a4). It should be noted, the critical time $t_c$ here reflects the time that the perturbations can be amplified. At this moment, the perturbations at the interface are still infinitesimal, which cannot be observed at the mesoscale. Hence, rather than the time when $\sigma_\omega$ turns positive, we define the crossover time based on a specific value of $A_\omega$. The time when the magnitude of $A_\omega$ reaches about 1μm is identified as the crossover time of the planar instability, shown by the purple curves in Fig. 4(b1)-(b4). Finally, the crossover times from the theoretical model are 0.63s, 0.67s, 0.69s, and 0.75s in Fig. 4, consisting of those from the PF simulations, 0.65s, 0.67s, 0.69s, and 0.76s in Fig. 3.

The consistencies between the two different models validate the accuracy of the simulations. Fig. 3 shows the crossover times of planar instability increase from $2.0\times10^{-9}$ to $6.0\times10^{-9}$. Meanwhile, the onset of the instability corresponds to different $V_I$, in Fig. 3(b), illustrating the velocity $V_I$ is not the criterion of the instability. By contrast, the onset of the instability corresponds to similar $c_0$ in Fig. 3(a). The results demonstrate the conclusion that the excess free energy at the interface and corresponding interfacial energy are the critical parameters of the instability [24]. Snapshots of the solute distribution at the crossover time are shown in Fig. 5(a1)-(a4). The larger $D_L$ corresponds to the larger diffusion length ahead of the interface, while the concentration ahead of the interface ($c_0$) is almost the same with different $D_L$.

The detailed evolution of the interfacial morphologies with different $D_L$ is shown in Fig. 5(b1)-(b4), where the time intervals between each two lines are the same ($\Delta t=0.043s$). On the one hand, the consuming time of the Planar-Cellular-Transition (PCT) increases from $2.0\times10^{-9}$ to $6.0\times10^{-9}$. Here the completion of the PCT is defined as the amplitude of the cellular becomes roughly comparable to its wavelength. On the other hand, the wavenumber of the cellular decreases from $2.0\times10^{-9}$ to $6.0\times10^{-9}$.

According to equation (3), the coefficient $D_L$ has a negative relation with the degree of dissipation.

The smaller $D_L$ means the larger degree of dissipation. Hence, the consuming time of the PCT increases from $2.0×10^{-9}$ to $6.0×10^{-9}$, shown in Fig. 5(b1)-(b4). Meanwhile, the solute atoms still accumulate at the interface during the PCT, till the appearing of the cellular. The longer time of the PCT means the longer time of solute accumulation. Hence, the peak of $c_0$ appears later and later from $2.0×10^{-9}$ to $6.0×10^{-9}$, in Fig. 3(a2), and the values of the peak increase from $2.0×10^{-9}$ to $6.0×10^{-9}$. For the velocity $V_I$, the larger $D_L$ means the lower degree of dissipation and the smaller $V_I$. In Fig. 3(b2), during the PCT stage, $V_I$ decreases from $2.0×10^{-9}$ to $6.0×10^{-9}$. The decreasing $V_I$ makes the interfacial curvature $\kappa$ decrease from $2.0×10^{-9}$ to $6.0×10^{-9}$. The smaller $\kappa$ means the larger wavelength and smaller wavenumber, i.e., the wavelength of the cellular increases from $2.0×10^{-9}$ to $6.0×10^{-9}$ while the wavenumber decreases, shown in Fig. 5(b1)-(b4).

In conclusion, at the planar growth stage, the larger $D_L$ corresponds to the smaller $c_0$ and the larger $V_I$. Different $V_I$ at the crossover times of planar instability illustrate the velocity $V_I$ is not the criterion of the instability. The decrease of the interfacial energy, induced by solute segregation at the S/L interface, can be regarded as the criterion of instability. During the PCT stage, the larger $D_L$ means the smaller degree of dissipation. As a result, the consuming time of the PCT increases from $2.0×10^{-9}$ to $6.0×10^{-9}$, while the tip velocity $V_I$ and the interfacial curvature $\kappa$ decreases from $2.0×10^{-9}$ to $6.0×10^{-9}$. Correspondingly, the wavelength of the cellular increases from $2.0×10^{-9}$ to $6.0×10^{-9}$ while the wavenumber decreases.

**2.2.2 The dendrite growth**

After the planar instability and the PCT stage, the dendrites (formed from the cellular) start to grow. At this stage, according to the rule of maximum surface energy, the crystal will seek to minimize the total energy by creating large curvature in the <100> direction, shown in Fig. 6(b1)-(b4). On the other hand, different $D_L$ make the evolutional characteristics differ from each other, including the concentration $c_0$ and tip velocity $V_I$ in Fig. 3(a2)-(b2), as well as the interfacial morphologies in Fig. 6.

After the appearing of the cellular, $c_0$ starts to decrease. As mentioned before, the tip curvature $\kappa$ and velocity $V_I$ decreases from $2.0×10^{-9}$ to $6.0×10^{-9}$. Correspondingly, the gradient $|\partial_n c|$ decreases from $2.0×10^{-9}$ to $6.0×10^{-9}$. Combined with equation (2), $V_I$ increases further and reaches its peak. The smaller $D_L$ corresponds to the larger $\kappa$, the smaller $c_0$, and the larger $|\partial_n c|$. Hence, the peak values of $V_I$ decrease from 2.0×10$^{-9}$ to 6.0×10$^{-9}$, shown in Fig. 3(b2). Meanwhile, the smaller $D_L$ means the larger degree of dissipation and non-equilibrium, leading to more non-equilibrium structures (sidebranches), shown in Fig. 6(b1)-(b4).

It should be noted, in Fig. 6 (a1)-(a4), the solute diffusion ahead of the interface is dominated by the magnitude of $D_L$, represented by the increasing diffusion length and the decreasing gradient of concentration from 2.0×10$^{-9}$ to 6.0×10$^{-9}$. By contrast, the solute diffusion between the dendrite trunks is affected by $D_L$ as well as the interfacial curvature, i.e., curvature-induced solute diffusion. During the directional solidification, due to the constrained growth conditions, the overall solute distribution is determined by the temperature. Hence, the gradients of solute concentration between the dendrite trunks show the same features.

In conclusion, the dendrites with smaller $D_L$ grow out more sidebranches. On the other hand, different $D_L$ makes the gradients of solute concentration ahead of the interface decrease from 2.0×10$^{-9}$ to 6.0×10$^{-9}$, while the gradients of concentration between the dendrite trunks show the same features.

**2.2.3 The steady-state growth**

As time goes further, solidification turns into the steady-state growth stage, shown by the stable curves in Fig. 3(a3)-(b3) and the morphological evolution in Fig. 7. At this stage, the dissipative solidification structures achieve a quasi-steady state after a period of self-organization. With different $D_L$, the overall propagation velocities of the S/L interface are the same and equal to the pulling speed $V_P$, shown in Fig. 3(b3). By contrast, due to the different $D_L$, the concentration at the interface $c_0$ decreases from 2.0×10$^{-9}$ to 6.0×10$^{-9}$, shown in Fig. 3(a3). In Fig. 7(a1)-(a4), on the one hand, the gradients of concentration ahead of the interface decrease from 2.0×10$^{-9}$ to 6.0×10$^{-9}$. On the other hand, under the constrained growth conditions of directional solidification, the overall distribution of solute distribution is determined by the temperature. The gradients of concentration between the dendrite trunks show the same features.

The morphological characteristics include the primary dendrite and sidebranches, determined by the primary dendrite arm space [25]. In Fig. 7, the space increases from 2.0×10$^{-9}$ to 6.0×10$^{-9}$, consisting of the rule of length scale selection, in equation (4), qualitatively [26].

$$\rho = \frac{1}{\sigma^*}\sqrt{d_0 \frac{D_L}{V_I}} \qquad (4)$$

where σ* is the selection constant, motivated by the value obtained for the minimum wavelength found in the stability analysis of planar growth. The larger $D_L$ corresponds to the larger tip radius ρ, the larger arm space, and the smaller number of primary dendrites.

The onset of sidebranches needs the primary arm space is greater than one critical value. Hence, we focus on Fig. 7(b3) and (b4) here, whose space is large enough. With the same primary space and tip velocity, the dendrites with smaller $D_L$ grow out more sidebranches. The result consists of the previous conclusion. The smaller $D_L$ means the larger degree of dissipation and non-equilibrium. Hence, more non-equilibrium structures (sidebranches) grow out.

To represent this phenomenon more clearly, the simulations with a given number of primary dendrites are performed (here the number is five). With different $D_L$ and the same number of primary dendrites, the steady-state growth is shown in Fig. 8. Although with the same primary dendrite arm space $\lambda_1$ and tip velocity $V_I$, the dendrites have different tip curvature κ. Specifically, κ decreases from $2.0\times10^{-9}$ to $6.0\times10^{-9}$. In addition, with the same $\lambda_1$ and $V_I$, the evolution of sidebranches shows differences. Fig. 8 shows both the number and amplitude of the sidebranches decrease from $2.0\times10^{-9}$ to $6.0\times10^{-9}$. The smaller $D_L$ means the larger degree of dissipation and non-equilibrium. Hence, even with the same $\lambda_1$ and $V_I$, the curvature κ and the number and amplitude of the sidebranches decrease from $2.0\times10^{-9}$ to $6.0\times10^{-9}$.

The phenomena reveal the following things. (1) With the same solidification parameters and material parameters, in Fig. 7 and Fig. 8, the evolution of solidification is quite different due to their different initial conditions. It consists of the fact that the initial conditions are deterministic for the following evolution in a dissipative system. (2) Comparing the evolution with different $D_L$ in Fig. 8, the interfacial morphologies do not have unique relations with the characteristic parameters. Due to different degrees of dissipation, even with the same arm space and tip velocity, the tip curvatures and the features of sidebranches are different.

In conclusion, at the steady-state stage, with different $D_L$, the overall propagation velocities of the S/L interface are the same and equal $V_P$. On the other hand, different $D_L$ means different degrees of dissipation. Although with the same tip velocity and primary arm space, the dendrites with different $D_L$ have different tip curvatures and different features of sidebranches.

The investigations in this paper indicate the importance of solute diffusion in alloy solidification. The dissipation at the interface is altered by the diffusion coefficient $D_L$. From the viewpoint of the whole domain, smaller $D_L$ corresponds to a higher degree of dissipation of the system, as well as more exchange of heat and mass with the environment. Hence, more S/L interfaces are formed during solidification. The areas of the interfaces decrease from $2.0\times10^{-9}$ to $6.0\times10^{-9}$, shown in Fig. 5(b1)-(b4), Fig. 6(b1)-(b4), Fig. 7 (b1)-(b4), and Fig. 8 (b1)-(b4).

In addition, the investigations explain the conclusion that the competitive influences of curvature and velocity depend strongly on the ratio $\tau/W^2$ in the governing equations [15]. In the current PF model, based on the expression of the kinetic coefficient β in equation (21) and the local equilibrium approximation (β=0), one can obtain the relation

$$\frac{\tau}{W^2} = a_2 \frac{\lambda}{D_L} \qquad (5)$$

According to equations (5) and (21), we know that $d_0$, $a_1$, $a_2$, W, and λ are constant during the PF simulation. Hence, the ratio $\tau/W^2$ is dominated by the diffusion coefficient $D_L$. Meanwhile, $D_L$ directly corresponds to the dissipation of the system. That is, the competitive influences of curvature and velocity (the ratio $\tau/W^2$) are because of the dissipation, resulting from the friction of atoms.

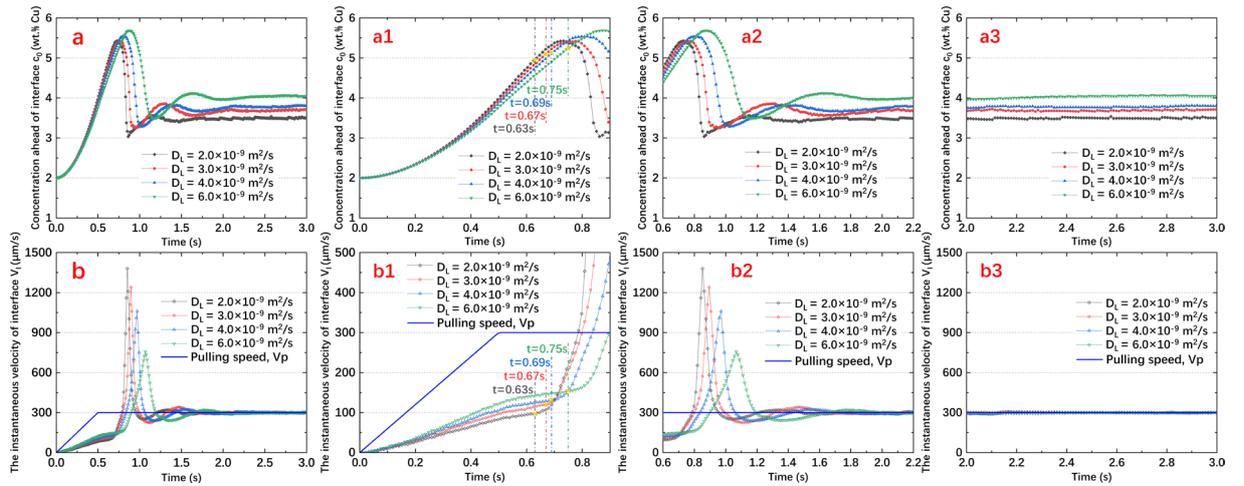

Fig. 3. The evolution of characteristic parameters with different diffusion coefficients $D_L$: (a) the concentration $c_0$ and (b) the instantaneous velocity $V_I$; (a1), (a2), and (a3) are the enlarged versions of (a); (b1), (b2), and (b3) are the enlarged versions of (b). (from the PF simulations)

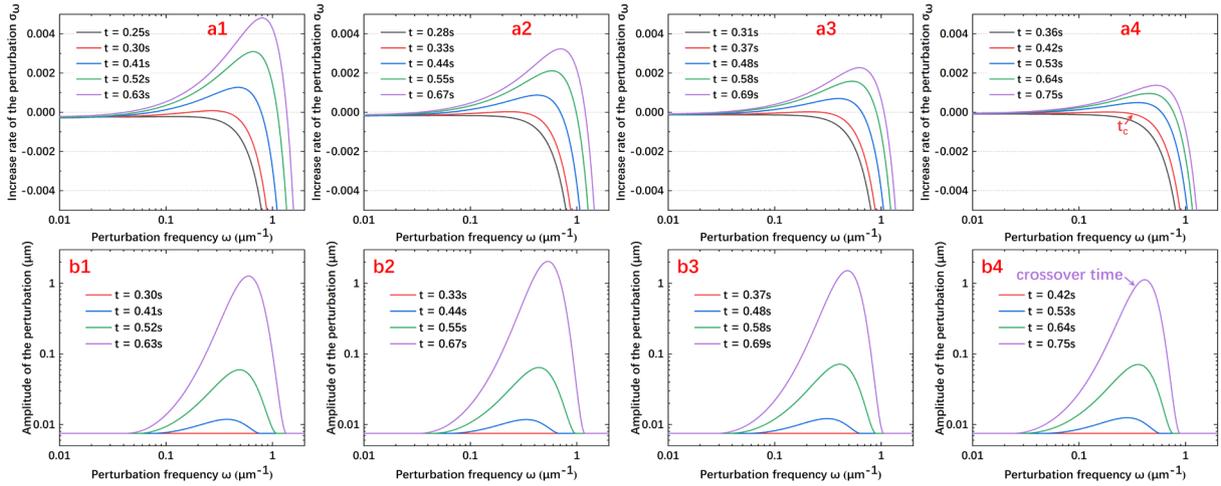

Fig. 4. Dynamical evolution of the instability predicted by the theoretical model with different solute diffusion coefficients $D_L$: (a) The increase rate of amplitude spectrum, (b) The evolution of the amplitude spectrum. (the diffusion coefficients are 2.0, 3.0, 4.0, and 6.0×10$^{-9}$ m$^2$/s, respectively)

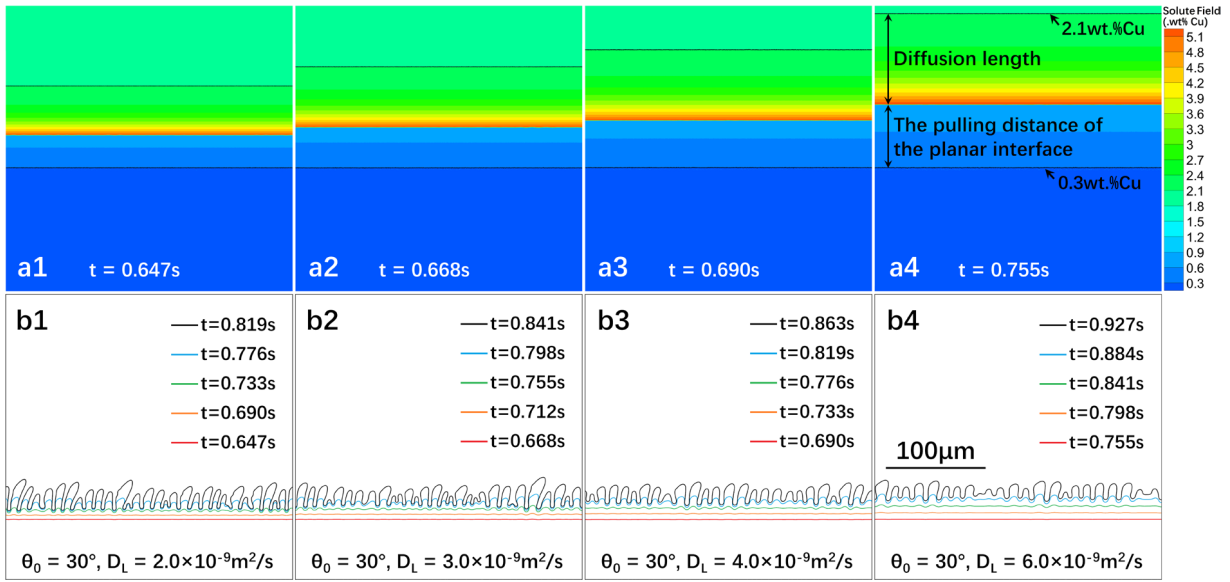

Fig. 5. (a) The solute distribution with different diffusion coefficient $D_L$ at the onset time of the planar instability; (b) the corresponding evolution of the interfacial morphologies with different $D_L$ at the Planar-Cellular-Transition stage. (from the PF simulations)

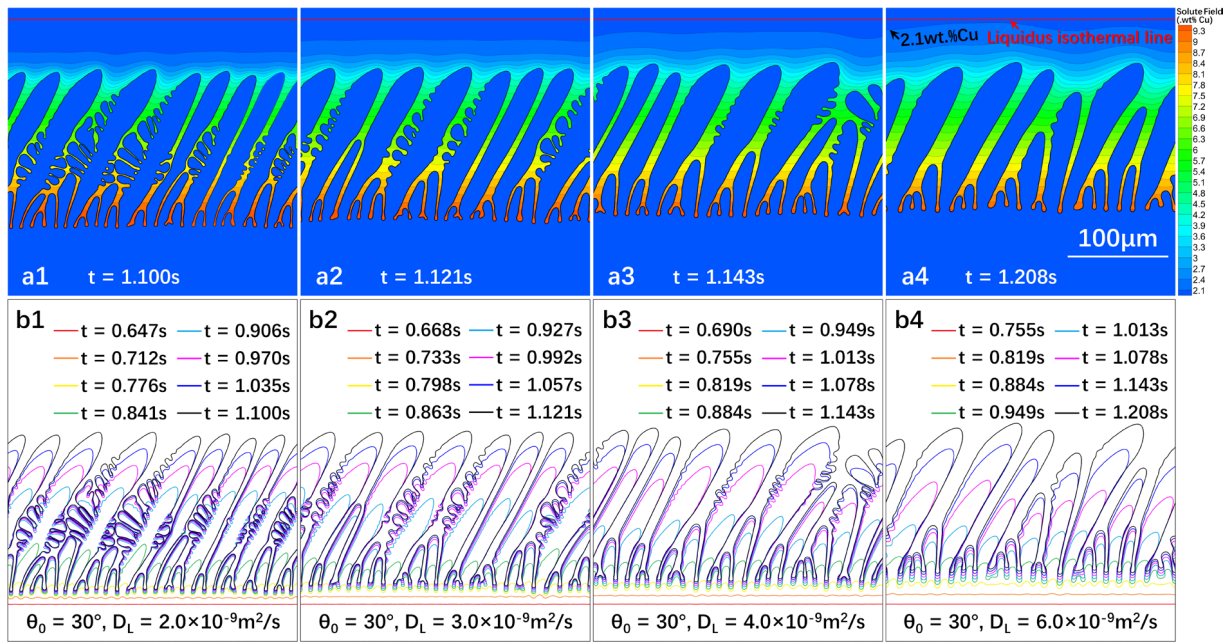

Fig. 6. (a) The solute distribution with different diffusion coefficient $D_L$ at the dendrite growth stage; (b) the corresponding evolution of the interfacial morphologies with different $D_L$ at this stage. (from the PF simulations)

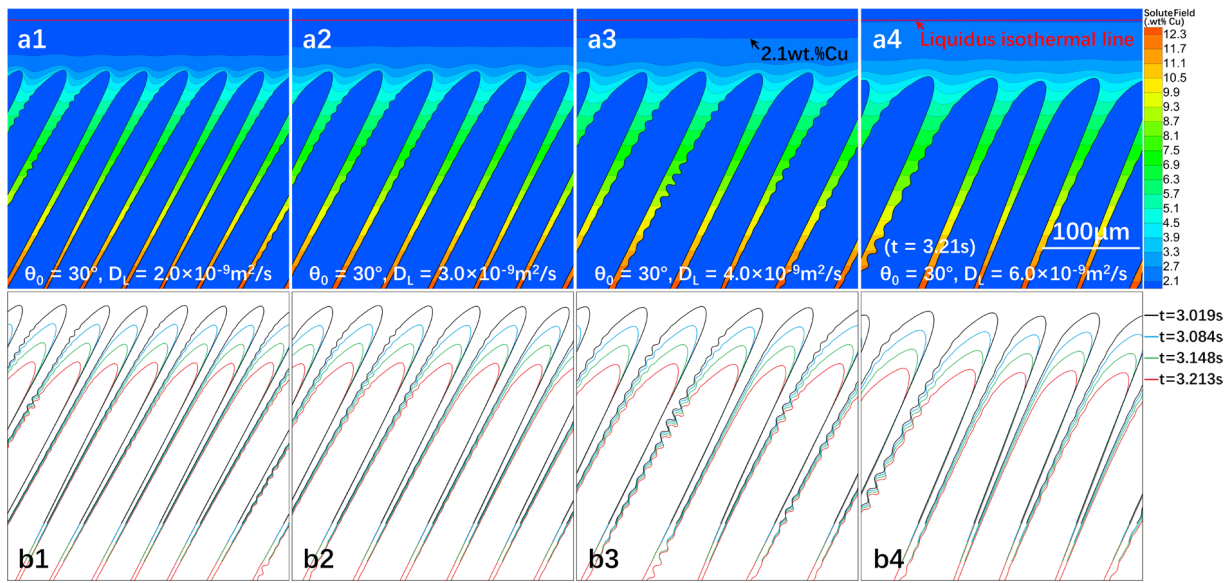

Fig. 7. (a) The solute distribution with different diffusion coefficient $D_L$ at the steady-state stage (t=3.21s); (b) the corresponding evolution of the interfacial morphologies with different $D_L$ at this stage. (from the PF simulations)

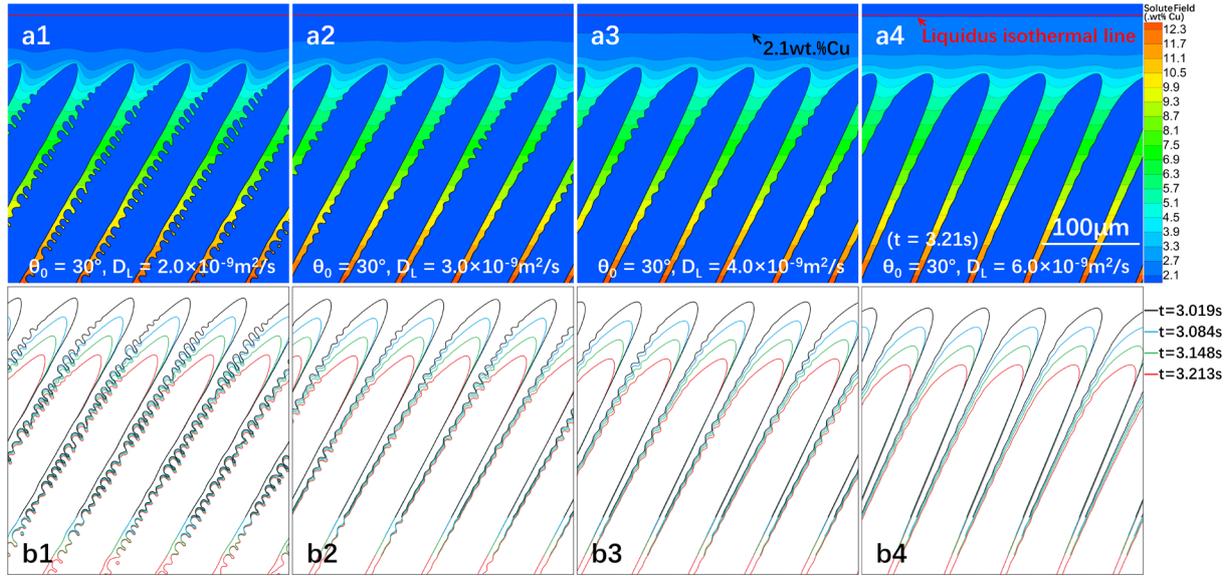

Fig. 8. (a) The solute distribution with the given number of dendrites and different diffusion coefficient $D_L$ at the steady-state stage (t=3.21s); (b) the corresponding evolution of the interfacial morphologies at this stage. (from the PF simulations)

### 3. Conclusions

This paper investigates the evolution in the whole domain during entire directional solidification via the quantitative PF model. The investigations demonstrate the dissipative features of solidification and indicate the important role of solute diffusion in alloy solidification. The dissipation at the interface is altered by the diffusion coefficient $D_L$. From the viewpoint of the whole domain, smaller $D_L$ corresponds to a higher degree of dissipation, as well as more exchange of heat and mass with the environment. Hence, more S/L interfaces are formed during solidification. In addition, the investigations explain the conclusion that the competitive influences of curvature and velocity depend strongly on the ratio $\tau/W^2$. The ratio $\tau/W^2$ is dominated by $D_L$, and $D_L$ corresponds to the dissipation. Hence, the competitive influences of tip curvature and velocity are because of the dissipation, resulting from the friction of atoms.

The discussion about the cause and influence factors of dissipation lays the fundament for investigating the competitive influences of the tip curvature and velocity on morphological evolution. The discussion about the effect of the diffusion coefficient on morphological evolution can provide theoretical bases for parameter optimization and material design, based on the magnitude of the diffusion coefficient, having great potential for engineering applications.

## 4. Models and methodology

For directional solidification, the so-called "frozen temperature approximation" is adopted

$$T(z,t) = T_0 + G\left(z - \int V_P(t)dt\right) \tag{6}$$

where $T_0$ is the melt temperature of the pure material, G is the thermal gradient, and $V_P$ is the pulling speed. The approximation is based on the assumptions: (1) The latent heat is ignored, i.e., the temperature field is undisturbed by the evolution of the Solid/Liquid (S/L) interface. It is essentially a statement concerning the relative magnitudes of the terms in the Stefan condition, $\rho_s L_f v^*_n \ll k_{s,l} \nabla T_{s,l} \cdot \mathbf{n}$ [27]. In the expression, $\rho_s$ is the density of the solid, $L_f$ is the latent heat, and $v^*_n$ is the rate of solidification. $k_{s,l}$ is the thermal conductivity, $\nabla T_{s,l}$ is the thermal gradient, and $\mathbf{n}$ is the normal direction of the interface. The rate of alloy solidification ($v^*_n$) is limited by solute diffusion. Since the coefficient of thermal conductivity is much larger than solute diffusion, the latent heat can be released quickly through heat conduction. Hence, the effect of latent heat on thermal transport could be ignored [28]. (2) There is no flow in the liquid, consistent with the assumption that the densities of the solid and liquid are equal.

It should be noted, the frozen temperature approximation is just for heat transport. The solidification process contains the latent heat, both in the theoretical model and the PF model.

### 4.1 Theoretical model

The theoretical model is based on the linear instability analysis under non-steady-state conditions, the detailed derivations can be found in references [7,29,30]. The following are the key equations.

The time-dependent concentration ahead of the interface $c_0$, position of the interface $z_0$, and diffusion length $l$ can be expressed as [7]

$$c_0(z_0,t) = \frac{2D_L c_\infty}{2D_L - V_I(1-k)l} \tag{7}$$

$$\frac{\partial z_0}{\partial t} = V_I - V_P(t) = \frac{2D_L(z_0 - z_\infty)}{l(1-k)z_0} - V_P(t) \tag{8}$$

$$\frac{\partial l}{\partial t} = \frac{4D_L(z_\infty - kz_0)}{l(1-k)z_0} - \frac{l}{z_0 - z_\infty}\frac{\partial z_0}{\partial t} \tag{9}$$

where $D_L$ is the diffusion coefficient in the liquid, $c_\infty$ is the average concentration, k is the solute partition coefficient, and $V_I$ is the instantaneous velocity of the interface. $z_\infty$ is the steady-state position of the planar interface with the relation of $z_\infty = -m \cdot c_\infty/G$, where G is the thermal gradient, and m is the slope of the liquidus line in the phase diagram.

Based on the time-dependent linear stability analysis and the assumption of an infinitesimal sinusoidal perturbation with the spacing frequency ω, the increase rate of perturbation amplitude can be given by

$$\sigma_\omega(t) = \left[ dA_\omega(t)/dt \right] / A_\omega(t) \tag{10}$$

where $A_\omega(t)$ is the amplitude. Combining the interface position and diffusion length in equations (8)-(9), the dispersion relation of the perturbation under transient conditions can be expressed by [30]

$$q_\omega \left\{ 1 + \frac{2\left[z_0(t) - z_\infty\right]}{l(t)} + \frac{\Gamma \omega^2}{G} \right\} = \\ \frac{V_I(t) - V_P(t)}{D_l} + \frac{2\left[z_0(t) - z_\infty\right]}{l(t)} \cdot \left[ \frac{V_I(t)}{D_l} + \frac{\sigma_\omega(t)}{V_I(t)} + \frac{1}{z_0(t)} + \frac{\Gamma \omega^2}{G \cdot z_0(t)} \right] \tag{11}$$

where $q_\omega$ is the inverse decay length of the concentration fluctuation at the interface along the z direction. Γ is the Gibbs-Thomson coefficient.

Based on equation (11), the time-dependent increase rate of the amplification $\sigma_\omega(t)$ can be obtained. Then according to the solution of equation (10), the time-dependent amplitude is

$$A_\omega(t) = A_\omega(0) \exp\left[ \int_{t_0}^{t} \sigma_\omega(t) dt \right] \tag{12}$$

where $t_0$ is the critical time when $\sigma_\omega$ changes from negative to positive. $A_\omega(0)$ refers to the initial amplitude of the infinitesimal fluctuation.

**4.2 Phase-Field model**

The following is a brief introduction to the PF model for alloy solidification. The driving force in this PF model is the difference in grand potential between phases [31,32]. The potential is related to the grand canonical ensemble of statistical mechanics, corresponding to an open system [33].

In the PF model, a scalar variable $\phi(\mathbf{r}, t)$ is introduced to identify the state of phases, where $\phi=+1$ reflects the solid, $\phi=-1$ reflects the liquid, and intermediate values of $\phi$ reflect the S/L interface. Since $\phi$ varies smoothly across the interface, the sharp interface becomes diffuse and the phases turn into a continuous field, i.e., phase field ϕ(**r**, t). For alloy solidification, the solute field c(**r**, t) is represented by the supersaturation field U(**r**, t):

$$U = \frac{1}{1-k}\left(\frac{2kc/c_\infty}{1+k-(1-k)\cdot\phi}-1\right) \quad (13)$$

Combined with the frozen temperature approximation, the governing equations of the phase field and supersaturation field are given by equations (14)-(15) [31,34]

$$a_s^2(\hat{n})\left[1-(1-k)\frac{z-\int V_P(t)dt}{l_T}\right]\frac{\partial\phi}{\partial t} =$$
$$\nabla\cdot\left[a_s^2(\hat{n})\vec{\nabla}\phi\right]-\partial_x\left(a_s(\hat{n})\cdot a_s'(\hat{n})\cdot\partial_y\phi\right)+\partial_y\left(a_s(\hat{n})\cdot a_s'(\hat{n})\cdot\partial_x\phi\right) \quad (14)$$
$$+\phi(1-\phi^2)-\lambda(1-\phi^2)^2\left[U+\frac{z-z_0-\int V_P(t)dt}{l_T}\right]$$

$$\left(\frac{1+k}{2}-\frac{1-k}{2}\phi\right)\frac{\partial U}{\partial t} = \nabla\cdot\left[D_L\cdot q(\phi)\cdot\vec{\nabla}U-\vec{j}_{at}\right]+\frac{1}{2}\left[1+(1-k)U\right]\frac{\partial\phi}{\partial t} \quad (15)$$

where,

$$l_T = \frac{\Delta T_0}{G(t)} = \frac{|m|c_\infty(1-k)}{kG(t)} \quad (16)$$

$$a_s(\hat{n}) \equiv a_s(\theta+\theta_0) = 1+\varepsilon_4\cos 4(\theta+\theta_0) \quad (17)$$

$$q(\phi) = \frac{1-\phi}{2} \quad (18)$$

$$\vec{j}_{at} = -\frac{1}{2\sqrt{2}}\left[1+(1-k)U\right]\frac{\partial\phi}{\partial t}\frac{\vec{\nabla}\phi}{|\vec{\nabla}\phi|} \quad (19)$$

In these equations, $l_T$ means the thermal length, where m is the slope of the liquidus line in the phase diagram. $a_s(n)$ is the four-fold anisotropy function in a 2D system, where $\varepsilon_4$ is the anisotropy strength, θ is the angle between the normal direction of the interface and the z-axis, and $\theta_0$ is the intersection angle between the Preferred Crystallographic Orientation (PCO) of the crystal and the Thermal Gradient Direction (TGD) (here the TGD is parallel to the z-axis). q(ϕ) is an interpolation function determining the varying solute diffusion coefficient across the domain. $\vec{j}_{at}$ is the Anti-Trapping

Current (ATC) term, where ∂ϕ/∂t means the rate of solidification, and ∇ϕ/|∇ϕ| is the unit length along the normal direction of the S/L interface.

It should be noted, equation (14) provides a clear connection between anisotropy of the microscopic and macroscopic levels. And it unifies the anisotropy in equilibrium and non-equilibrium conditions [15,35]. In equation (15), the ATC term counterbalances the trapping current associated with the jump of chemical potential across the interface and modifies the mass conservation condition at the interface. Specifically, the concentrations on both sides of the interface vary with velocity, they do not satisfy the partition relation (out of equilibrium). The jump of supersaturation can be interpreted as resulting from a finite interface mobility that leads to interface dissipation [34].

In performing quantitative simulations, it is essential to obtain the precise relations between the calculation parameters in the equations and the real physical qualities. The asymptotic analysis can achieve this goal. Specifically, the perturbation analyses have been performed on each scale, including the inner scale (interface) and the outer scale (sharp-interface problem), then the two expansions are matched [36]. In this way, the calculation parameters in the PF equations and the physical qualities could be linked by

$$d_0 = a_1 \frac{W}{\lambda} \tag{20}$$

$$\beta = a_1 \left( \frac{\tau}{\lambda W} - a_2 \frac{W}{D_L} \right) \tag{21}$$

In these equations, W and τ are the interface width and relaxation time, which are the length scale and time scale, respectively. $a_1 = 5\sqrt{2}/8$ and $a_2 = 47/75$. $d_0 = \Gamma/|m|(1-k)(c_\infty/k)$ is the chemical capillary length, where $\Gamma = \gamma_{sl} T_f/(\rho_s L_f)$ is the Gibbs-Thomson coefficient, $\gamma_{sl}$ is S/L interfacial energy, $T_f$ is the melting point of pure solvent, and $L_f$ is the latent heat. Equation (20) describes the Gibbs-Thomson effect, where λ is the coupling constant. Equation (21) expresses the relation between the kinetic coefficient β and other parameters. It contains two terms. The first term describes the dissipation due to a homogeneous undercooling of the interface. The second term represents the inhomogeneity of the temperature field inside the interface [37]. Due to the local equilibrium approximation, β is set to be zero in the simulations.

### 4.3 Computational procedure

During the computation, the material parameters of Al-2.0wt.%Cu are shown in Table 1 [38,39].

Table 1. The material parameters of Al-2.0wt.%Cu for simulations [38,39]

| Symbol | Value | Unit |
| --- | --- | --- |
| Liquidus temperature, $T_L$ | 927.8 | K |
| Solidus temperature, $T_S$ | 896.8 | K |
| Diffusion coefficient in liquid phase, $D_L$ | $3.0 \times 10^{-9}$ | m²/s |
| Equilibrium partition coefficient, k | 0.14 | / |
| Alloy composition, $c_\infty$ | 2.0 | wt.% |
| Liquidus slope, m | -2.6 | K/wt.% |
| Gibbs-Thomson coefficient, $\Gamma$ | $2.4 \times 10^{-7}$ | K·m |
| Anisotropic strength of surface energy, $\varepsilon_4$ | 0.01 | / |

When solving equations (7)-(9) numerically, for a small time interval $\Delta t$, the relations in equations (22)-(23) can be regarded as the initial conditions.

$$l \approx \left(\frac{8D_L \cdot \Delta t}{3}\right)^{1/2} \tag{22}$$

$$z_0 = z_\infty - V_P(t) \cdot \Delta t + \frac{V_P(t)\sqrt{2D_L}}{\sqrt{3} \cdot z_\infty (1-k)} (\Delta t)^{3/2} \tag{23}$$

When solving equation (11), the dispersion relation, the increase rate $\sigma_\omega$ satisfies equation(24) [7]

$$\sigma_\omega = D_l \left(q_\omega^2 - \omega^2\right) - q_\omega V_I \tag{24}$$

Setting $\sigma_\omega = 0$, the solution of equation (24) is

$$q_\omega = \frac{V_I}{2D_l} + \sqrt{\omega^2 + \left(\frac{V_I}{2D_l}\right)^2} \tag{25}$$

Eliminating $q_\omega$ from these equations and inserting the values of $z_0$ and $l$, the time-dependent spectrum of the increase rate $\sigma_\omega(t)$ can be obtained. Taking $\sigma_\omega(t)$ to equation (12), the time-dependent amplitude can be obtained. In equation (12), $A_\omega(0)$ reflects the initial amplitude. Based on the approximation of equilibrium fluctuation spectrum, $A_\omega(0)$ equals the capillary length $d_0$ [30].

When solving equations (14)-(15), the PF governing equations, the most important parameter is the interface width W [16]. The simulation accuracy increases as W decreases, while the computational cost increases dramatically as W decreases. Performing the thin interface limitation, the magnitude of W just needs to be one order of the magnitude smaller than the characteristic length scale of the structure [31,40]. Since the characteristic length of alloy solidification is $L_C \sim \sqrt{d_0}*D_L/V_{tip}$ [27], W was set to be 0.15μm. In the computation, the periodic boundary conditions were loaded for both the phase field and supersaturation field. The time step size was chosen below the threshold of numerical instability for the diffusion equations, i.e., $\Delta t<(\Delta x)^2/(4D_L)$. Finally, when $D_L$ is $3.0\times10^{-9} m^2/s$, this paper used the fixed grid size $\Delta x=0.8W$ and the time step size $\Delta t=0.013\tau$.

Moreover, a fluctuating current $J_U$ is introduced to the diffusion equation to consider the infinitesimal perturbation of thermal noise on the S/L interface. By using the Euler explicit time scheme

$$U^{t+\Delta t} = U^t + \Delta t \left( \partial_t U - \vec{\nabla} \cdot \vec{J}_U \right) \tag{26}$$

The components of $J_U$ are random variables obeying a Gaussian distribution, which has the maximum entropy relative to other probability distributions [41]

$$\left\langle J_U^m(\vec{r},\vec{t}) J_U^n(\vec{r}\,',\vec{t}\,') \right\rangle = 2D_L q(\phi) F_U^0 \delta_{mn} \delta(\vec{r}-\vec{r}\,')\delta(t-t') \tag{27}$$

In equation (27), the constant noise magnitude $F_U^0$ reflects the magnitude of $F_U$ for a reference planar interface at temperature $T_0$, defined as [42,43]

$$F_U^0 = \frac{k v_0}{(1-k)^2 N_A c_\infty} \tag{28}$$

where $v_0$ is the molar volume of the solute atom, and $N_A$ is the Avogadro constant. According to the Clausius-Clapeyron relation

$$\frac{|m|}{1-k} = \frac{k_B T_0^2}{\Delta h} \tag{29}$$

where $\Delta h$ is the latent heat, and $k_B$ is the Boltzmann constant. The constant noise amplitude becomes

$$F_U^0 = \frac{k}{|m|c_\infty(1-k)} \frac{k_B T_0^2}{L} \tag{30}$$

Finally, the program codes of the theoretical model and PF model were written in C++. The explicit Finite Difference Method (FDM) was used when solving the governing equations, and the Message Passing Interface (MPI) parallelization was adopted for improving the computational efficiency.

## Data and code availability

The sharing of the data and code in this paper is available upon request.

## Declarations

The author declared that there is no conflict of interest.